\documentclass[11pt]{article}

\usepackage[utf8]{inputenc}
\usepackage[T1]{fontenc}
\usepackage{textcomp}
\usepackage[margin=1in]{geometry}
\usepackage{amsmath}
\usepackage{txfonts}  
\usepackage{graphicx}
\usepackage{booktabs}
\usepackage{array}
\usepackage{xcolor}
\usepackage{microtype}
\usepackage[numbers,sort&compress]{natbib}
\usepackage{authblk}
\usepackage[colorlinks=true,linkcolor=blue!60!black,citecolor=blue!60!black,urlcolor=blue!60!black]{hyperref}
\usepackage{url}

\let\origunderscore\_
\renewcommand{\_}{\origunderscore\penalty300\relax}
\title{io\_uring in Oracle Database: A Hybrid Storage I/O Architecture at Production Scale}

\author{Rajarshi Chowdhury}
\author{Akshay Shah}
\author{Margaret Susairaj}
\author{Ayush Agrawal}
\affil{Oracle America Inc., Redwood Shores, CA 94065, USA\\
\texttt{\char123 rajarshi.chowdhury, akshay.shah, margaret.susairaj, ayush.ag.agrawal\char125@oracle.com}}

\date{}

\begin{document}

\maketitle

\begin{abstract}
We describe the integration of io\_uring into Oracle Database's storage layer and the architectural decisions required to deploy it in a production multi-process RDBMS. Our design uses per-process ring contexts that eliminate inter-process synchronization, a shared buffer registration mechanism now part of the mainline Linux kernel, and a transparent fallback to libaio on error. Evaluation on an internal development build of Oracle Database 26ai shows that io\_uring's benefits concentrate on asynchronous batched I/O paths: on a mixed OLTP workload (TPC-C), io\_uring delivers identical throughput while reducing server CPU utilization by 1.2 percentage points through more efficient background writes; on analytical queries (TPC-H), CPU per query drops by 8.5\% (geometric mean). Isolating the write path alone shows 29\% lower CPU per write and 34\% higher throughput. Synchronous read paths---the dominant I/O in OLTP---show no improvement and even increased CPU usage for very large IO sizes. These results motivate a hybrid I/O architecture that retains pread/pwrite for synchronous operations, adopts io\_uring for asynchronous batched I/O, and falls back transparently to libaio---a selective strategy that may also be relevant to other database systems facing the same integration question.
\end{abstract}

\section{Introduction}
\label{sec:intro}

Modern NVMe storage devices can sustain millions of IOPS and single-digit microsecond latencies, yet conventional Linux I/O interfaces struggle to keep pace~\cite{haas2023nvme, leis2024cloudnative, bjorling2013linux}. The traditional kernel I/O stack imposes system call overhead, context switches, and memory copies that consume a significant fraction of CPU cycles---cycles that database systems need for query processing, buffer management, and concurrency control. As storage hardware continues to improve, this software overhead becomes an increasingly dominant bottleneck~\cite{zhou2025oltp, yang2012don, caulfield2010moneta}.

The Linux \texttt{io\_uring} interface~\cite{axboe2019iouring, axboe2022iouring_evolution}, introduced in kernel 5.1, addresses these inefficiencies through shared ring buffers between user space and kernel, batched submission of multiple I/O operations via a single system call, and asynchronous completion notification. Microbenchmarks show that naive adoption yields only modest gains over \texttt{libaio}, but architecture-aware integration---exploiting batching and asynchronous execution---can more than double throughput~\cite{didona2022understanding, ren2023performance}. These results have generated considerable interest in the database community. Jasny et al.~\cite{jasny2026iouring} provide the most thorough analysis to date, evaluating \texttt{io\_uring} across database buffer management workloads (random and sequential page access, TPC-H-style scans) and a distributed shuffle operator, establishing that batching and asynchronous execution are prerequisites for meaningful benefit. Their study uses a single-process, fiber-based architecture---a setting that does not encounter the multi-process deployment challenges of a production RDBMS.

However, there is a gap between microbenchmark promise and production reality. Deploying \texttt{io\_uring} in a production multi-process RDBMS such as Oracle Database raises challenges on two fronts. First, a production database issues fundamentally diverse I/O patterns: synchronous single-block reads on index lookups, batched asynchronous writes from background buffer pool flushers, large sequential reads for parallel table scans, and latency-critical redo log writes requiring durability. Not all of these benefit from \texttt{io\_uring}. Second, Oracle Database uses a multi-process architecture where hundreds of foreground server processes and dozens of background processes (database buffer writers, log writers, parallel query slaves) independently issue I/O against a shared buffer pool and shared data files. This architecture differs fundamentally from the single-process, thread-per-ring model assumed by most \texttt{io\_uring} studies~\cite{jasny2026iouring, didona2022understanding, ren2023performance}.

In this paper, we describe the integration of \texttt{io\_uring} into Oracle Database's storage layer, addressing these challenges and evaluating the results across read, write, and analytical scan workloads. Our contributions are:

\begin{enumerate}
    \item \textbf{A hybrid I/O architecture} that retains \texttt{pread}/\texttt{pwrite} for synchronous paths, adopts \texttt{io\_uring} for asynchronous batched I/O, and falls back transparently to \texttt{libaio} on error. Our results suggest that this selective adoption---rather than wholesale replacement---is a sound strategy for multi-process databases.
    
    \item \textbf{A shared buffer registration mechanism} initiated by Oracle and now part of the mainline Linux kernel~\cite{clone_buffers_kernel612}, enabling Oracle's shared buffer pool to be registered once and reused across hundreds of per-process \texttt{io\_uring} rings without redundant memory overhead.
    
    \item \textbf{A systematic characterization of I/O path suitability} that identifies which database I/O patterns benefit from \texttt{io\_uring} (batched asynchronous) and which do not (synchronous, small-batch), leading to a per-process ring architecture where each process or thread owns a private \texttt{io\_uring} context, eliminating inter-process synchronization.
\end{enumerate}

We evaluate the design using microbenchmarks and on an internal development build of Oracle Database 26ai across read-intensive, write-intensive, and parallel scan workloads.

\section{Background}
\label{sec:background}
The \texttt{io\_uring} interface~\cite{axboe2019iouring} uses two ring buffers shared between user space and kernel: a \emph{Submission Queue} (SQ) for I/O requests and a \emph{Completion Queue} (CQ) for results. This design enables batched submission---multiple operations via a single \texttt{io\_uring\_enter} system call---and lock-free completion retrieval by reading the CQ directly from user space. Table~\ref{tab:iouring-comparison} compares the interfaces. We refer the reader to Axboe~\cite{axboe2019iouring} and Jasny et al.~\cite{jasny2026iouring} for a thorough description of \texttt{io\_uring}.

\begin{table}[htbp]
\centering
\caption{System call comparison across I/O interfaces.}
\label{tab:iouring-comparison}
\small
\begin{tabular}{lccc}
\hline
\textbf{Operation} & \textbf{Synchronous} & \textbf{libaio} & \textbf{io\_uring} \\
\hline
Setup        & ---               & \texttt{io\_setup}       & \texttt{io\_uring\_setup} \\
             &                   &                          & \texttt{io\_uring\_register} \\
Submit       & \texttt{pread64}  & \texttt{io\_submit}      & \texttt{io\_uring\_enter} \\
             & \texttt{pwrite64} &                          & \\
Reap         & (implicit)        & \texttt{io\_getevents}   & (CQ peek, no syscall) \\
\hline
Syscalls per & 1 per I/O         & 2 per batch              & 1 per batch \\
batch        &                   & (submit + reap)          & (submit; reap is free) \\
\hline
\end{tabular}
\end{table}

The key advantage for database workloads is syscall consolidation: \texttt{libaio} requires two syscalls per batch (\texttt{io\_submit} + \texttt{io\_getevents}), while \texttt{io\_uring} combines submission and optional completion waiting into one call, with zero-syscall reaping from user space.

\begin{figure}[t]
\centering
\includegraphics[width=\textwidth]{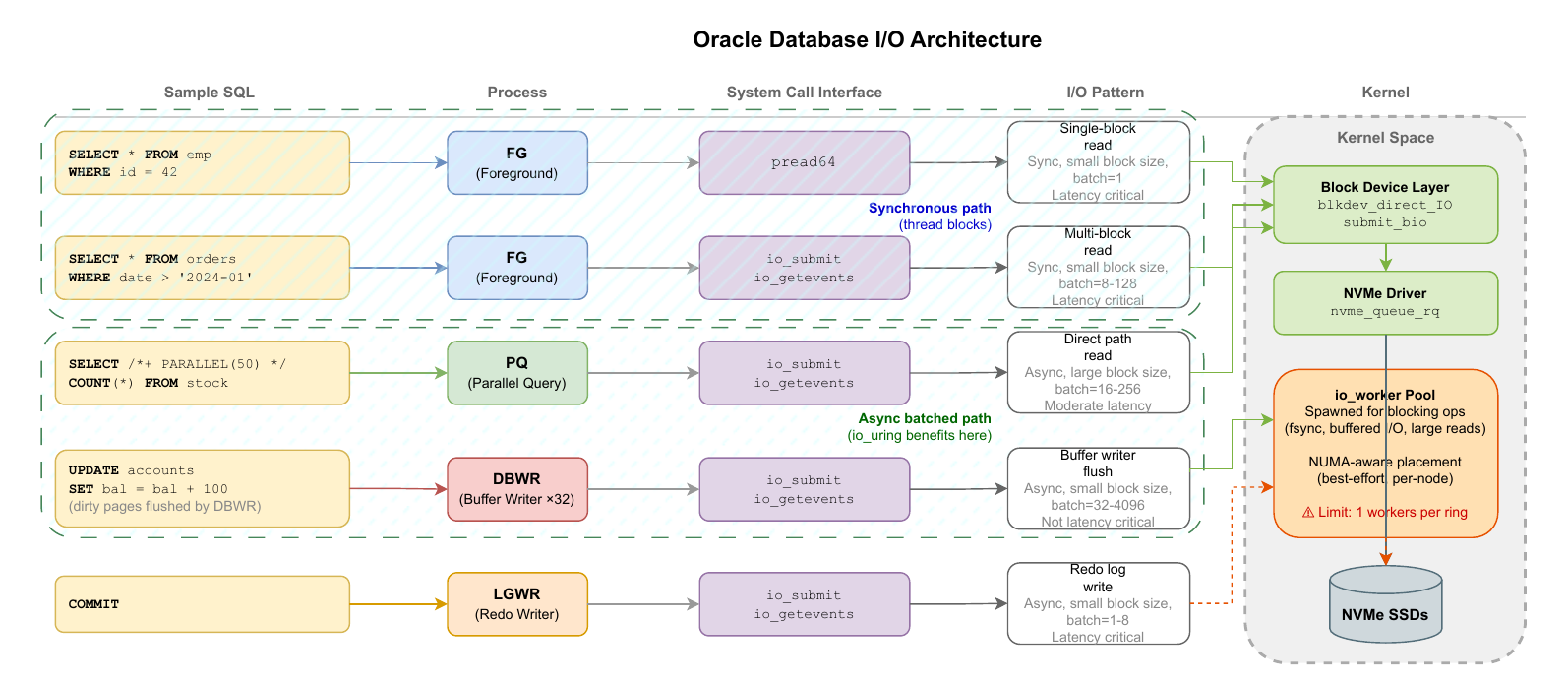}
\caption{Oracle Database I/O architecture (pre-\texttt{io\_uring}). Foreground (FG) processes use synchronous \texttt{pread64} for single-block reads and \texttt{libaio} for multi-block reads. Parallel query (PQ) slaves and buffer writers (DBWR) use \texttt{libaio} for asynchronous batched I/O. Green-highlighted rows indicate the asynchronous batched paths where \texttt{io\_uring} provides benefit.}
\label{fig:oracle-io-arch}
\end{figure}

\texttt{io\_uring} supports three execution modes~\cite{axboe2019iouring, jasny2026iouring}: \emph{interrupt-driven} (default), where completions arrive via interrupts; \emph{IOPOLL}, where the application polls NVMe completion queues directly, trading CPU for lower latency; and \emph{SQPOLL}, where a dedicated kernel thread polls the SQ, eliminating submission syscalls at the cost of a dedicated core. Several registration mechanisms further reduce per-I/O overhead: file descriptor registration avoids per Submission Queue Entry (SQE) reference lookups, ring fd registration reduces \texttt{io\_uring\_enter} overhead, buffer registration pins user-space buffers for direct DMA~\cite{corbet2020io_uring_buffers}, and the single-issuer flag enables internal optimizations when only one task submits to a ring~\cite{axboe2019iouring, didona2022understanding}. These features become important in our implementation (Section~\ref{sec:architecture}).

\subsection{Oracle Database I/O Architecture}
\label{sec:bg-oracle}

Oracle Database employs a multi-process architecture~\cite{oracleioarch, oracle_concepts} in which each client connection is served by a dedicated \emph{Foreground} (FG) server process, and shared \emph{background} processes handle buffer pool management, logging, and other system tasks. The key background processes relevant to I/O are: the \emph{Database Buffer Writer} (DBWR), which flushes dirty pages from the shared buffer cache to disk; the \emph{Log Writer} (LGWR), which writes redo log records to persistent storage to ensure transaction durability; and \emph{Parallel Query} (PQ) worker processes, which execute parallel scan operations. All of these processes share access to the \emph{System Global Area} (SGA), a large shared memory region that contains the buffer cache, redo log buffer, and other shared data structures. These processes issue I/O through distinct code paths depending on the operation type, as illustrated in Figure~\ref{fig:oracle-io-arch}.

\textbf{Single-block reads} occur when a foreground process accesses a block not in the buffer cache, issuing a \texttt{pread64} that blocks until completion. In Oracle's wait model, the FG process posts a \emph{blocking wait} and is descheduled by the OS until the I/O completes. This is the most frequent I/O in OLTP workloads, with no batching opportunity.

\textbf{Multi-block reads} arise during table scans and index range scans. The database submits a batch of requests via \texttt{libaio} (\texttt{io\_submit} + \texttt{io\_getevents}), but the foreground thread waits for all reads before processing, issuing an immediate reap after each submit. The FG process posts a blocking wait until all blocks in the batch arrive. This immediate-reap pattern limits the opportunity for syscall amortization, as the submit-and-reap cycle behaves similarly across interfaces.

\textbf{Direct path reads} are issued by PQ workers during full table scans, bypassing the buffer cache and reading directly into process-private memory. Each PQ worker issues large asynchronous reads (typically 1\,MB) with up to 256 outstanding I/Os, making this the most throughput-oriented read pattern. PQ workers use a \emph{polling wait} model: after submitting a batch, they continue processing previously completed I/Os and periodically poll the completion queue for new results, overlapping computation with I/O.

\textbf{Buffer writer flushes} are performed by DBWR processes, which write dirty pages from the buffer cache in large asynchronous batches (up to 4096 writes) via \texttt{libaio}. Oracle Database typically runs multiple DBWR processes (e.g., 32) to sustain high write throughput. Each DBWR uses a polling wait model similar to PQ workers: it submits a batch of writes, then polls for completions while preparing the next batch, maintaining a deep pipeline of outstanding I/Os.

\textbf{Redo log writes} are performed by the LGWR process, which flushes log records to persistent storage using \texttt{libaio}. This I/O is latency-critical: committing transactions post a blocking wait until LGWR confirms their log records are durable. Redo writes are typically small batches (fewer than 8 I/Os) and are reaped immediately after submission, limiting the opportunity for syscall amortization.

The distinction between blocking and polling waits is important for understanding io\_uring's applicability. Blocking waits (single-block reads, multi-block reads, Redo Writes) cause the process to sleep until I/O completes---there is no opportunity to overlap work, so io\_uring's asynchronous completion model provides no advantage. Polling waits (DBWR flushes, PQ direct-path reads) allow the process to continue useful work while I/O is in flight, making them natural candidates for io\_uring's batched submission and zero-syscall completion reaping.

The key observation is that Oracle Database's I/O paths are already differentiated by synchrony, batch size, and latency sensitivity. Any \texttt{io\_uring} integration must respect these distinctions rather than apply a uniform replacement strategy. Furthermore, the multi-process, multi-threaded model means that each process and thread independently manages its own I/O context---unlike single-process, single-thread engines where a central event loop can coordinate all I/O through a shared ring~\cite{jasny2026iouring}.

\section{Characterizing io\_uring Benefits}
\label{sec:iopatterns}

Before committing to an io\_uring integration we needed to determine which
I/O patterns described above benefit from the new interface.  Table~\ref{tab:iopatterns}
summarizes the result: only batched asynchronous paths gain a meaningful
advantage.  This section presents the microbenchmark evidence behind that
conclusion.

All experiments use a single NVMe device with six progressively optimized configurations, from a \texttt{libaio} baseline through batched io\_uring with registration and polling.  Each thread simulates 10--100\,$\mu$s of CPU work between I/O rounds to model production conditions where the database spends cycles on query processing between physical reads. The think time bounds per-thread throughput at \textasciitilde250\,K IOPS regardless of device speed; aggregate throughput at higher thread counts is much higher (Table~\ref{tab:reg_breakdown}). All microbenchmark numbers reported are the mean of 100 runs; variance was below 1\% across all configurations.  The benchmark source code is available in the accompanying artifact.

\begin{table}[htbp]
\centering
\caption{Oracle I/O patterns and io\_uring applicability.}
\label{tab:iopatterns}
\small
\begin{tabular}{lcc>{\raggedright\arraybackslash}p{6.2cm}}
\toprule
\textbf{I/O Pattern} & \textbf{Sync?} & \textbf{Benefit} & \textbf{Rationale} \\
\midrule
Single-block read   & Yes & None  & Already one syscall (\texttt{pread}) \\
Multi-block read    & No  & Negligible & Immediate reap; FG waits for all reads \\
Direct-path read    & No  & High & Async batched submit amortizes syscalls \\
Buffer writer flush & No  & High & 32--128 pages per flush \\
Redo log write      & No & Negligible & Small batches ($<$8), immediate reap \\
\bottomrule
\end{tabular}
\end{table}

\subsection{Syscall Amortization}
\label{sec:syscall}

io\_uring's advantage does not come from faster individual I/O operations---a
single NVMe read takes the same time regardless of the submission
interface~\cite{haas2023nvme}.  The advantage comes from \emph{syscall amortization}: filling
$N$ submission-queue entries in userspace and submitting them with a single
\texttt{io\_uring\_enter}, then reaping completions from the completion
queue without a separate syscall~\cite{axboe2019iouring, soares2010flexsc}.

With \texttt{libaio}, every I/O batch requires two syscalls
(\texttt{io\_submit} + \texttt{io\_getevents}).  \texttt{io\_uring}
consolidates these: a single \texttt{io\_uring\_enter} submits all
pending SQEs, and completions are reaped from the shared CQ in user
space without a syscall.  On a per-batch basis, this is a 2$\times$
reduction in syscall count (2 syscalls to 1).  The practical benefit
compounds with batch depth: at a batch size of~32, \texttt{io\_uring}
issues one syscall for 32~I/Os and reaps for free, yielding
${\sim}0.03$ syscalls per I/O; \texttt{libaio} issues two syscalls for
the same 32~I/Os, yielding ${\sim}0.06$ syscalls per I/O.  The
difference is modest per I/O, but at hundreds of thousands of IOPS the
aggregate CPU savings become significant, as we show below.

\begin{figure}[t]
\centering
\includegraphics[width=\textwidth]{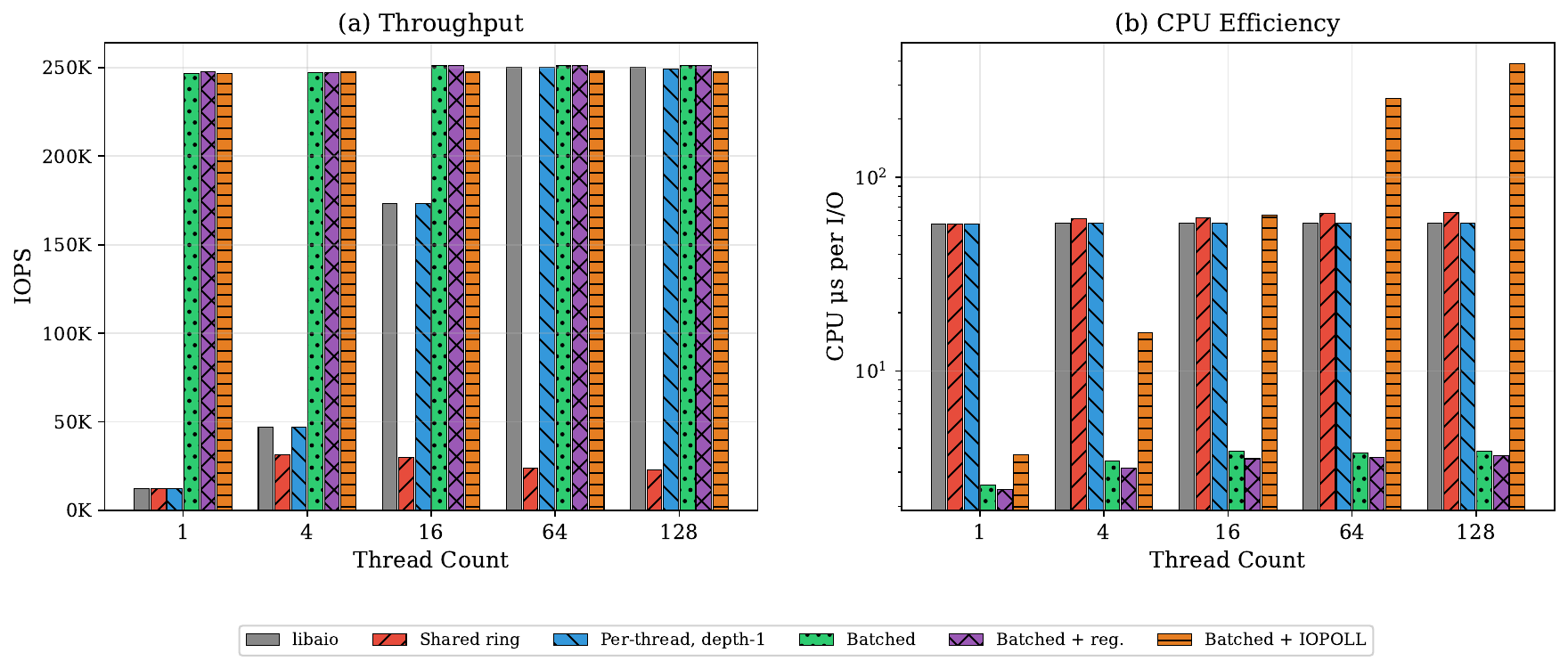}
\caption{Microbenchmark results across six io\_uring configurations and
five concurrency levels on a single NVMe device.
\textbf{(a)}~Throughput: the batched pipeline saturates the device
(${\sim}250$\,K IOPS) from a single thread; the shared-ring anti-pattern
collapses under concurrency.
\textbf{(b)}~CPU efficiency (log scale): batched pipelines consume
${\sim}3.5\,\mu$s of CPU per I/O; IOPOLL's per-thread busy-polling
scales to 385\,$\mu$s at 128 threads---100$\times$ more than
the interrupt-driven batched pipeline at the same concurrency.}
\label{fig:bench}
\end{figure}

This amortization is directly visible in Figure~\ref{fig:bench}a.
The batched configurations reach the per-thread think-time limit at
\textasciitilde247\,K~IOPS from a single thread, whereas the depth-1
configurations require 64 threads to reach the same throughput because
each I/O incurs a full kernel round-trip before the next can be submitted.
Figure~\ref{fig:bench}b confirms the CPU consequence: the batched pipeline
consumes only 2.5--3.9\,$\mu$s of CPU per I/O, compared to
\textasciitilde58\,$\mu$s for depth-1 configurations whose threads
busy-spin during the simulated think time between synchronous I/Os.

\subsection{Synchronous Paths: No Benefit}

Single-block reads---the most common I/O in OLTP---are inherently
synchronous: the server process issues one \texttt{pread} and blocks until
the page is in the buffer cache.  Replacing \texttt{pread} with an
io\_uring submit-and-wait adds ring management overhead without reducing
syscalls.  Figure~\ref{fig:bench} confirms this: at every thread count,
per-thread depth-1 io\_uring matches \texttt{libaio} on both throughput
and CPU efficiency.

Table~\ref{tab:sync_read} further validates this across block sizes from 4\,KB to 1\,MB. At every block size, \texttt{pread} and io\_uring deliver essentially identical IOPS and average latency. At 512\,KB and 1\,MB, io\_uring's CPU cost increases due to kernel worker thread offload for large I/Os that cannot complete inline, confirming that synchronous reads should remain on pread. The 512\,KB--1\,MB rows are included for completeness rather than to characterize a common path: synchronous foreground reads in Oracle Database OLTP workloads are dominated by 4--16\,KB single-block reads (the database block size, typically 8\,KB), and large reads at this scale are issued asynchronously by parallel-query and direct-path code paths rather than synchronously by foreground sessions. The large-block rows therefore show that even where sync io\_uring would be most expensive, that combination does not arise in typical foreground OLTP access patterns.

\begin{table}[htbp]
\centering
\caption{Synchronous single-block read: \texttt{pread} vs.\ io\_uring submit-and-wait, single thread, 30\,s per block size.}
\label{tab:sync_read}
\small
\begin{tabular}{r rrr rrr}
\toprule
 & \multicolumn{3}{c}{\textbf{pread}} & \multicolumn{3}{c}{\textbf{io\_uring}} \\
\cmidrule(lr){2-4} \cmidrule(lr){5-7}
\textbf{Block} & \textbf{IOPS} & \textbf{Avg} & \textbf{CPU} & \textbf{IOPS} & \textbf{Avg} & \textbf{CPU} \\
\textbf{Size}  &               & \textbf{($\mu$s)} & \textbf{($\mu$s/IO)} &       & \textbf{($\mu$s)} & \textbf{($\mu$s/IO)} \\
\midrule
  4\,KB  & 49\,878 & 20.0 &  1.23 & 49\,088 & 20.3 &  1.41 \\
  8\,KB  & 37\,360 & 26.7 &  1.30 & 36\,940 & 27.0 &  1.46 \\
 16\,KB  & 30\,105 & 33.2 &  1.38 & 29\,732 & 33.6 &  1.55 \\
 32\,KB  & 20\,818 & 48.0 &  1.61 & 20\,728 & 48.2 &  1.73 \\
 64\,KB  & 13\,347 & 74.9 &  1.99 & 13\,326 & 75.0 &  2.08 \\
128\,KB  &  7\,567 & 132.1 & 2.75 &  7\,561 & 132.2 & 2.71 \\
256\,KB  &  4\,067 & 245.8 & 4.05 &  4\,058 & 246.4 & 4.00 \\
512\,KB  &  2\,119 & 471.8 & 7.33 &  2\,078 & 481.2 & 16.18 \\
  1\,MB  &  1\,677 & 596.4 & 13.44 & 1\,648 & 606.7 & 27.00 \\
\bottomrule
\end{tabular}
\end{table}

A natural follow-up question is whether the synchronous path could be unified with the asynchronous one by splitting each large foreground read into multiple smaller chunks (e.g., a 1\,MB read split into four 256\,KB requests) and relying on the device's native command queuing to recombine the underlying access. Two chunking strategies exist within io\_uring. \emph{Linked SQEs} (\texttt{IOSQE\_IO\_LINK} / \texttt{IOSQE\_IO\_HARDLINK}) sequence dependent operations such that each linked SQE only starts after the previous completes; this is designed for dependent chains (e.g., read a header, then read the body at the disclosed offset) and would \emph{serialize} the chunks, driving total latency to roughly $N$~$\times$~per-chunk time. \emph{Unlinked batched submission}---$N$ independent chunk SQEs issued in one \texttt{io\_uring\_enter} with \texttt{wait\_nr}~=~$N$---could in principle recover the io-wq offload penalty that single-submit io\_uring incurs at $\geq$512\,KB (Table~\ref{tab:sync_read}), since $\leq$256\,KB chunks complete inline; but the Linux block layer already splits large pread requests into BIOs that exploit device-level NCQ parallelism, so the available headroom is narrow, and the chunked path adds SQE/CQE management plus user-space recombination cost that pread does not pay. More fundamentally, neither strategy changes the property that ultimately matters: a synchronous wait removes the one mechanism io\_uring is designed to exploit---overlap of computation with in-flight I/O---regardless of how the request is decomposed. We therefore keep all single-block and multi-block foreground reads on pread/libaio.

\subsection{Ring Topology: Shared vs.\ Per-Process}

An early design question was whether DBWR processes or a bunch of FG processes or the PQ processes should share a single
io\_uring instance (protected by a mutex) or each maintain a private ring.
The shared-ring approach simplifies resource management but serializes
every submit and reap operation.

Table~\ref{tab:p99} quantifies the cost.  At 4 threads, the shared ring's
p99 latency is already 1.8$\times$ that of the per-thread configuration.
At 128~threads the gap widens to 26$\times$ (22\,ms vs.\ 850\,$\mu$s),
and throughput collapses to 23\,K~IOPS---less than one-tenth of the
250\,K~IOPS achieved by per-thread rings (Figure~\ref{fig:bench}a).
The mutex serializes not only the \texttt{io\_uring\_enter} syscall but
also SQE fill and CQ drain, creating a convoy effect.

\begin{table}[htbp]
\centering
\caption{p99 latency ($\mu$s) for depth-1 configurations. The shared-ring
mutex creates a convoy effect at high concurrency.}
\label{tab:p99}
\small
\begin{tabular}{lrrrrr}
\toprule
\textbf{Configuration} & \textbf{1T} & \textbf{4T} & \textbf{16T} & \textbf{64T} & \textbf{128T} \\
\midrule
libaio & 100 & 100 & 100 & 360 & 850 \\
io\_uring Shared ring & 100 & 180 & 1.4\,ms & 10.4\,ms & 22.1\,ms \\
io\_uring Per-thread, depth-1 & 100 & 100 & 100 & 360 & 850 \\
\bottomrule
\end{tabular}
\end{table}

\subsection{Polling Modes: Infeasible at Production Scale}
\label{sec:polling}
io\_uring supports two polling modes beyond the default interrupt-driven completion: \texttt{IOPOLL} and \texttt{SQPOLL}. We found neither to be a good fit for our design.

\textbf{IOPOLL} replaces
interrupt-driven completion with busy-polling inside
\texttt{io\_uring\_enter}.  The kernel spins on the NVMe completion queue
until entries appear, eliminating interrupt latency at the cost of CPU
cycles.

At low concurrency the tradeoff is modest: with one thread, IOPOLL adds
roughly 1\,$\mu$s of CPU per I/O compared to the interrupt-driven
batched pipeline (3.7 vs.\ 2.5\,$\mu$s).  But the cost diverges at
scale because each thread independently busy-polls its completion queue.
Figure~\ref{fig:bench}b shows that at 128~threads, IOPOLL consumes
385\,$\mu$s of CPU per I/O while the interrupt-driven batched pipeline
consumes 3.9\,$\mu$s at the same concurrency---a 100$\times$ gap.
Both costs scale with thread count, but IOPOLL's busy-wait dominates
because polling time is proportional to the number of active threads,
whereas interrupt-driven completion incurs only a fixed per-interrupt cost
shared across the batch.
Database servers are commonly run with substantial CPU headroom;
dedicating two orders of magnitude more CPU to I/O completion would leave
no headroom for query processing.

\textbf{SQPOLL} spawns a dedicated kernel thread that continuously polls the submission queue, eliminating submission syscalls entirely. On modern kernels ($\geq 5.13$) it is available to unprivileged processes, but it remains impractical for Oracle Database for two reasons. First, SQPOLL dedicates one kernel thread per ring to busy-polling the SQ. With hundreds of per-process rings in a production instance, this would consume hundreds of CPU cores solely for submission polling. Second, the SQPOLL thread runs at the kernel's scheduling priority and sits outside any cgroup the database manages, making its CPU consumption invisible to Oracle Resource Manager's per-session quotas. We rely on syscall amortization via batched io\_uring\_enter rather than elimination via SQPOLL, accepting the per-batch syscall as the cost of staying inside Oracle's resource accounting boundary.

On older kernels ($\leq 5.12$) SQPOLL did require \texttt{cap\_sys\_nice}, but the deeper objections above apply regardless of the privilege model and remain decisive on modern kernels where no capability is required.

\subsection{Shared Buffer Registration}
\label{sec:bufreg}

io\_uring allows processes to \emph{register} I/O buffers with the kernel
via \texttt{IORING\_REGISTER\_BUFFERS}.  Registration pins the pages and
pre-computes the kernel mapping, eliminating per-I/O \texttt{get\_user\_pages}
calls.  However, Oracle Database operates as a multi-process architecture
with a shared buffer cache that can reach tens of
gigabytes~\cite{oracle_concepts}.  If each of the $N$ DBWR processes independently registers the
same buffer cache region, the kernel must call \texttt{get\_user\_pages} $N$ times
on the same physical pages---a cost that grows linearly with both buffer
size and process count.

Table~\ref{tab:bufreg_size} shows registration time for a single process
as a function of buffer size.  The cost is dominated by
\texttt{get\_user\_pages} and scales linearly: a 64\,GB buffer takes
590\,ms, pinning 16.7 million pages.  At Oracle's typical buffer cache sizes
(16--64\,GB), a single registration completes in 145--590\,ms---acceptable
as a one-time startup cost.

\begin{table}[htbp]
\centering
\caption{Buffer registration time vs.\ buffer size (single process).
The cost is dominated by \texttt{get\_user\_pages()} and scales linearly
with the number of pages pinned.}
\label{tab:bufreg_size}
\small
\begin{tabular}{rrrr}
\toprule
\textbf{Buffer} & \textbf{Reg.\ Time} & \textbf{Rate} & \textbf{Pages} \\
\textbf{Size}   & \textbf{(ms)}       & \textbf{(GB/s)} & \textbf{Pinned} \\
\midrule
  1\,GB &   19 &  51 &     262\,K \\
  4\,GB &   48 &  83 &   1\,049\,K \\
  8\,GB &   81 &  99 &   2\,097\,K \\
 16\,GB &  146 & 110 &   4\,194\,K \\
 32\,GB &  272 & 118 &   8\,389\,K \\
 64\,GB &  590 & 108 &  16\,777\,K \\
\bottomrule
\end{tabular}
\end{table}

The problem emerges at scale.  Table~\ref{tab:bufreg_procs} shows
wall-clock time when multiple processes each register the same 16\,GB
buffer independently.  With 200 processes the total time reaches 2.0\,s,
and the slowest individual process waits over 2.0\,s due to contention on
the kernel's \texttt{mmap\_sem} during concurrent \texttt{get\_user\_pages}
calls.  In contrast, a shared registration model---where one manager
process registers the buffer and others attach to the existing
registration---completes in 7.5\,ms for 200 processes, because no
additional page pinning is required.  This 270$\times$ gap motivated the
shared buffer registration mechanism described in
Section~\ref{sec:shared-bufreg}.

\begin{table}[htbp]
\centering
\caption{Buffer registration scaling with multiple processes (16\,GB buffer).
Per-process registration grows linearly; shared registration stays constant.}
\label{tab:bufreg_procs}
\small
\begin{tabular}{rrrr}
\toprule
 & \multicolumn{2}{c}{\textbf{Per-Process}} & \textbf{Shared} \\
\cmidrule(lr){2-3} \cmidrule(lr){4-4}
\textbf{Procs} & \textbf{Wall (ms)} & \textbf{Avg (ms)} & \textbf{Wall (ms)} \\
\midrule
    1 &     615 &  615 &    0.8 \\
    5 &   1\,258 &  976 &    0.4 \\
   10 &   1\,265 & 1\,116 &    0.7 \\
   25 &   1\,323 & 1\,141 &    1.7 \\
   50 &   1\,330 & 1\,209 &    2.7 \\
  100 &   1\,831 & 1\,537 &    5.4 \\
  200 &   2\,044 & 1\,916 &    7.5 \\
\bottomrule
\end{tabular}
\end{table}

Buffer registration is the dominant of three registration optimizations enabled at ring setup, the others being file descriptor (FD) and ring-fd registration (Section~\ref{sec:custom-api}). To guide systems with smaller buffer pools on which to port first, we measured the incremental contribution of each on the same batched io\_uring pipeline used in Figure~\ref{fig:bench}, scaled to 16 threads (8\,KB random reads, batch 32--128, 60\,s, single NVMe; same 10--100\,$\mu$s think time between batches as Figure~\ref{fig:bench}). Table~\ref{tab:reg_breakdown} reports the result. Ring-fd registration alone is within run-to-run noise; adding file-fd shaves 1.4\% off CPU per I/O; adding buffer registration accounts for essentially the entire benefit, dropping CPU per I/O from 3.66 to 2.58~$\mu$s, a 29.5\% reduction. IOPS is unchanged across all four configurations because the device is at saturation at this thread count. Systems that sidestep buffer registration (e.g.\ fork-based engines that inherit page mappings from the postmaster) capture only a small fraction of the available CPU saving from io\_uring's registration features.

\begin{table}[htbp]
\centering
\caption{Per-optimization io\_uring registration breakdown on the batched async pipeline (16 threads, 8\,KB reads, 60\,s, single NVMe). Deltas vs.\ no-registration baseline.}
\label{tab:reg_breakdown}
\small
\begin{tabular}{lrrrr}
\toprule
\textbf{Configuration} & \textbf{IOPS} & \textbf{$\Delta$ IOPS} & \textbf{CPU/IO ($\mu$s)} & \textbf{$\Delta$ CPU} \\
\midrule
No registration & 1\,701\,K & ---       & 3.66 & ---       \\
$+$ ring-fd     & 1\,702\,K & $+$0.1\% & 3.66 & 0.0\%     \\
$+$ file-fd     & 1\,706\,K & $+$0.3\% & 3.61 & $-$1.4\% \\
$+$ buffer      & 1\,705\,K & $+$0.2\% & 2.58 & $-$29.5\% \\
\bottomrule
\end{tabular}
\end{table}

\subsection{Redo Log Writes: No Benefit}
\label{sec:redo-micro}

Redo log writes are latency-critical but use small batches (typically 1--8 I/Os) with immediate reap after each submit. To confirm that io\_uring provides no meaningful advantage for this pattern, we ran a redo-like microbenchmark: 4 threads issuing 512\,B--4\,KB sequential writes with batch sizes of 1--8 and immediate reap, for 60 seconds on a single NVMe device.

\begin{table}[htbp]
\centering
\caption{Redo-like microbenchmark: 512\,B--4\,KB writes, batch 1--8, immediate reap, 4 threads, 60\,s.}
\label{tab:redo_micro}
\small
\begin{tabular}{lrr}
\toprule
\textbf{Metric} & \textbf{libaio} & \textbf{io\_uring} \\
\midrule
Total I/Os          & 10\,980\,K & 11\,008\,K \\
IOPS                & 183\,003   & 183\,467   \\
I/Os per submit     & 4.5        & 4.5        \\
CPU per I/O ($\mu$s) & 13.32     & 13.47      \\
Avg latency ($\mu$s) & 9.6       & 9.6        \\
p99 latency ($\mu$s) & 110       & 110        \\
\bottomrule
\end{tabular}
\end{table}

Table~\ref{tab:redo_micro} shows that the two interfaces are indistinguishable on every metric: throughput, CPU efficiency, and tail latency are all within noise. With an average batch size of 4.5 and immediate reap, io\_uring's syscall consolidation advantage is negligible---the submit-then-immediately-reap cycle requires one syscall under both interfaces (io\_uring reaps from the CQ in user space, but the immediate wait effectively makes it synchronous). We route redo log writes through io\_uring nonetheless, to maintain a single asynchronous I/O code path across all background processes, simplifying the implementation without introducing any regression.

\subsection{Sensitivity to Block Size and Device Count}
\label{sec:sensitivity}

The microbenchmarks so far hold block size and device count fixed.
To check that the io\_uring CPU advantage on the async batched path
is robust across operating points, we ran the same
batched-libaio-vs-batched-io\_uring comparison over a $4 \times 6$
grid of $\{1, 2, 4, 8\}$ device counts and
$\{$4\,KB, 8\,KB, 16\,KB, 64\,KB, 256\,KB, 1\,MB$\}$ block sizes,
16 worker threads throughout, 30\,s per data point. Both interfaces
use a single combined submit-and-reap call per cycle (\texttt{io\_submit}
+ \texttt{io\_getevents} for libaio, \texttt{io\_uring\_enter} with
\texttt{IORING\_ENTER\_GETEVENTS} for io\_uring) so that batch depths
are comparable across interfaces. A 0.5--1.5\,$\mu$s per-IO busy spin
models the per-block CPU work that real database operators (row
processing, checksum, LRU update) perform between completions.

\begin{figure}[htbp]
\centering
\includegraphics[width=0.6\textwidth]{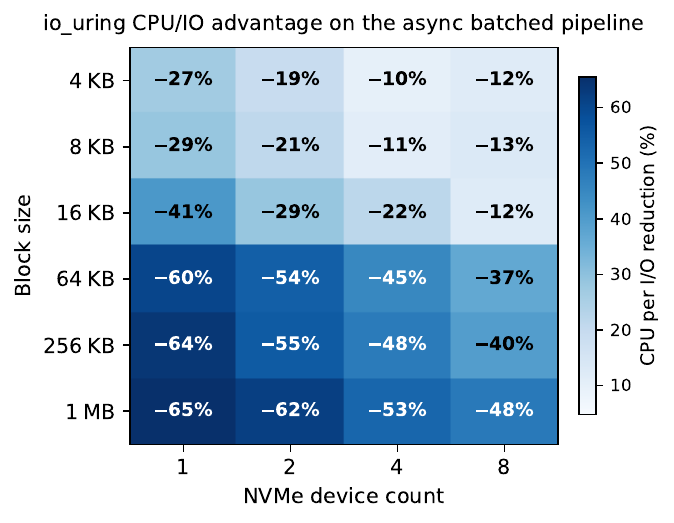}
\caption{CPU per I/O reduction for io\_uring (with full registration,
io-wq capped at one worker per ring) versus libaio on the async
batched pipeline, swept over $\{1, 2, 4, 8\}$ NVMe devices and
$\{$4\,KB$\,$--$\,$1\,MB$\}$ block sizes. All cells favor io\_uring;
darker blue indicates larger savings. Geometric mean across the
24 cells: $-39\%$ CPU per I/O.}
\label{fig:sensitivity_heatmap}
\end{figure}

Figure~\ref{fig:sensitivity_heatmap} reports the resulting CPU per I/O
reduction across the grid. The advantage is monotonic in block size
and present at every operating point we measured: from $-10\%$ in the
small-block, multi-device regime up to $-65\%$ at 1\,MB on a single
device, with a geometric mean of $-39\%$ over the 24 cells. Two factors
compound. First, syscall amortization: io\_uring's combined enter call
submits and reaps in one syscall, while libaio requires
\texttt{io\_submit} plus \texttt{io\_getevents}; the saving is fixed
per call and therefore a smaller fraction of total work at large
block sizes. Second, registration: at large block sizes the kernel's
\texttt{get\_user\_pages} cost grows linearly with the I/O size, and
io\_uring's pre-pinned buffers eliminate it. The right-hand column
(8 dev) shows the smaller savings characteristic of CPU-bound regimes
where syscall overhead is already a small fraction of total work; the
left-hand column (1 dev) shows the larger savings characteristic of
device-saturated regimes where the fixed per-IO kernel work dominates.
Throughput tells a complementary story (not shown in the figure): in
the small-block, multi-device regime where CPU was the bottleneck,
io\_uring delivers up to $+15\%$ IOPS at the same CPU; everywhere
else IOPS match the libaio baseline and the win is realized as freed
CPU rather than higher throughput.

\section{Architecture and Implementation}
\label{sec:architecture}

\begin{figure}[t]
\centering
\includegraphics[width=0.6\textwidth]{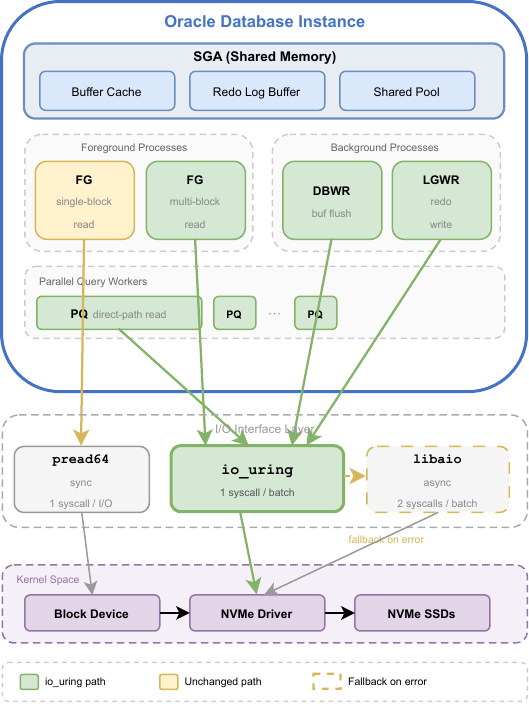}
\caption{Hybrid I/O architecture. Asynchronous paths (DBWR flushes, direct-path reads, and redo log writes) use \texttt{io\_uring} by default with transparent \texttt{libaio} fallback. Synchronous reads remain on \texttt{pread64}.}
\label{fig:hybrid-arch}
\end{figure}

Guided by the analysis in Section~\ref{sec:iopatterns}, we adopt a selective hybrid architecture. All asynchronous paths---DBWR buffer writer flushes, parallel query direct path reads, multi-block reads, and redo log writes---are routed through \texttt{io\_uring}. Multi-block reads and redo log writes see negligible performance difference due to their immediate-reap patterns, but are included for implementation consistency: maintaining a single asynchronous I/O interface simplifies the code path. Synchronous single-block reads remain on \texttt{pread64} and \texttt{libaio} serves as a transparent fallback for all \texttt{io\_uring} paths on error (Section~\ref{sec:fallback}). Figure~\ref{fig:hybrid-arch} illustrates this routing.

Each process that uses \texttt{io\_uring} owns a private ring instance, avoiding the shared-ring serialization quantified in Section~\ref{sec:iopatterns}. Oracle Database can operate in a dedicated-process model (one process per connection), a multi-threaded model (a shared pool of threads serving multiple connections), or a combination of both. Regardless of the execution model, each process or thread that issues asynchronous I/O creates its own ring on first use (128 to 4996 entries depending on type of process, ${\sim}$80---240\,KB of kernel memory) and uses it exclusively until exit. This per-process/per-thread ownership eliminates inter-process synchronization on the I/O path. All evaluations in this paper use the dedicated-process model, a widely used configuration. Ring ownership semantics are the same in the multi-threaded model. Each thread that issues I/O creates and exclusively owns a private ring, so the absence of shared-ring contention does not depend on the execution model. The mechanical difference is that threads in a multi-threaded process share a single virtual address space. Registered buffers (Section~\ref{sec:shared-bufreg}) and registered file descriptors are therefore visible to all threads in that process, while the ring file descriptors themselves remain per-thread. The kernel-side data structures touched on the I/O path (SQ tail, CQ head, registered buffer table) are the same in both models, so the per-I/O cost is identical. With 200 concurrent processes the aggregate cost is roughly 24\,MB---negligible relative to a typical buffer cache of 16--128\,GB.

\subsection{Custom API Layer}
\label{sec:custom-api}

The standard \texttt{liburing} library~\cite{liburing} provides a convenient API for io\_uring and is well-suited for most applications. For Oracle Database, however, we implement a thin custom layer that interfaces directly with the \texttt{io\_uring} system calls and shared memory rings. This allows us to integrate tightly with Oracle's existing I/O tracking, error handling, and resource lifecycle without adapting to \texttt{liburing}'s abstractions. The layer exposes four functions: \emph{setup}, \emph{prepare}, \emph{submit}, and \emph{reap}.

\textbf{Setup} creates a per-process \texttt{io\_uring} context. It invokes \texttt{io\_uring\_setup} with \texttt{IORING\_SETUP\_CQSIZE} (sizing the CQ equal to the SQ) and \texttt{IORING\_SETUP\_CLAMP}, maps the SQ/CQ rings and SQE array via \texttt{mmap} (using \texttt{IORING\_FEAT\_SINGLE\_MMAP} where available), stores pointers to ring control fields for direct user-space access, and pre-populates the SQ array so that slot $i$ maps to SQE $i$. It sets \texttt{FD\_CLOEXEC} on the ring file descriptor to prevent leakage across \texttt{fork}/\texttt{exec} boundaries. Since each ring has exactly one submitter, we set \texttt{IORING\_SETUP\_SINGLE\_ISSUER} to enable kernel-side optimizations (gracefully skipped on pre-6.0 kernels).

Sizing the CQ equal to the SQ is intended to prevent completions from overflowing the CQ ring, because each submitted SQE produces at most one CQE. Oracle Database's upper I/O layer independently tracks the number of submitted, completed, and outstanding I/Os per ring, and never submits more SQEs than the ring depth allows. This design is intended to avoid CQ overflow, and with it the complexity of the kernel's overflow recovery mechanism.

A subtle but important resource control is the \texttt{io-wq} worker pool. When the kernel cannot complete an I/O inline (e.g., for buffered I/O or operations that block internally), it offloads the work to a per-ring pool of \texttt{io-wq} kernel threads~\cite{axboe2019iouring}. By default, the kernel may spawn multiple workers per ring. In our implementation, however, nearly all I/O is direct (\texttt{O\_DIRECT}) and completes in-band---the NVMe driver processes the submission synchronously within the \texttt{io\_uring\_enter} syscall and the completion arrives via interrupt without involving any worker thread. We therefore limit the \texttt{io-wq} pool to one worker per ring ~\cite{axboe_iowq_limit}. This matters at scale: a large database instance may run hundreds of processes, each owning a private ring. Without the limit, the kernel could spawn thousands of idle worker threads, consuming kernel memory and scheduling resources for a code path that is almost never exercised.
Setup also applies a stack of registration optimizations:

\begin{itemize}
\item \emph{Ring file descriptor registration}: eliminates the \texttt{\_\_fdget()} overhead on each io\_uring\_enter.
\item \emph{File descriptor registration}: all data file descriptors open at ring creation time are registered, allowing SQEs to reference files by index and avoiding per-SQE \texttt{fget}/\texttt{fput} overhead. When new data files are opened subsequently, they are added to the ring's registered file set.
\item \emph{Buffer registration}: handled through a shared registration mechanism described in Section~\ref{sec:shared-bufreg}.
\end{itemize}

Each registration step is individually optional: if any fails (unsupported kernel, insufficient locked memory), the ring operates without that optimization and I/O proceeds correctly, albeit with higher per-operation overhead.

\textbf{Prepare} populates a single SQE without submitting it. It indexes into the SQE array at the current SQ tail (masked by ring size), fills the SQE fields, stores a caller-provided pointer in \texttt{user\_data} for completion identification, and advances the tail. The upper layer calls prepare repeatedly---once per dirty page or read extent---building up a batch before submission.

\textbf{Submit} pushes all pending SQEs to the kernel in one  \texttt{io\_uring\_enter} call. It peeks the SQ to count waiting entries (tail minus kernel-visible head) and submits them, returning immediately without waiting for completions.

\textbf{Reap} retrieves completed I/Os. It first peeks the CQ by comparing head and tail pointers---a pure user-space operation. If the minimum requested completions are not yet available and the caller requests blocking, reap calls \texttt{io\_uring\_enter} with \texttt{IORING\_ENTER\_GETEVENTS}, simultaneously submitting any pending SQEs. It also supports a non-blocking mode that returns whatever completions are available. A typical DBWR cycle calls prepare for each dirty page, submit once for the batch, then reap periodically---in the common case, completions have already arrived and the CQ peek succeeds without a syscall. This reduces the per-batch syscall count from two (\texttt{io\_submit} + \texttt{io\_getevents}) to one, with zero-syscall reaping when completions outpace consumption.

\subsection{Shared Buffer Registration}
\label{sec:shared-bufreg}

As shown in Section~\ref{sec:bufreg}, per-ring buffer registration is prohibitively expensive at Oracle's scale. To address this, Oracle initiated work on shared buffer registration for the Linux kernel~\cite{sharedreg_patch}: Bijan Mottahedeh proposed the initial RFC patch series introducing \texttt{IORING\_SETUP\_SHARE\_BUF} and \texttt{IORING\_SETUP\_ATTACH\_BUF} flags, and contributed the foundational design across multiple revisions (v1--v5). This work also generalized the existing file registration infrastructure into a unified resource registration framework, enabling buffer updates and sharing to reuse the same reference-counting and lifecycle management code paths. The io\_uring community subsequently evolved the concept into the \texttt{IORING\_REGISTER\_CLONE\_BUFFERS} operation, which was merged in Linux kernel 6.12~\cite{clone_buffers_kernel612}.

The mainline mechanism is file-descriptor (FD) based: a source ring registers buffers via the standard \texttt{IORING\_REGISTER\_BUFFERS} path, pinning the pages and creating the kernel-side buffer table. A destination ring then issues \texttt{IORING\_REGISTER\_CLONE\_BUFFERS} with the source ring's file descriptor. The kernel clones the existing buffer table into the destination ring without calling \texttt{get\_user\_pages} again, completing in constant time regardless of buffer size. Kernel 6.13 extended this with support for partial-range cloning and in-place replacement of existing registrations.

In our implementation, a single background process creates its ring and registers the entire buffer cache once---a one-time cost of ${\sim}$146\,ms for a 16\,GB buffer cache. Each DBWR, PQ worker, FG or other process that subsequently creates a ring clones the buffer registration from the manager ring's file descriptor, attaching to the existing registration in under 0.1\,ms per process. For 200 processes, the total registration time drops from 2.0\,s (per-process) to 7.5\,ms (shared)---a 270$\times$ improvement (Table~\ref{tab:bufreg_procs}). The registration remains valid as long as at least one ring holds a reference; when the last ring is torn down, the kernel unpins the pages.

The manager process is not a single point of failure. The kernel's \texttt{IORING\_REGISTER\_CLONE\_BUFFERS} operation \emph{clones} the buffer table into the destination ring rather than \emph{linking} it back to the source, so each destination ring holds its own independent reference on the pinned pages. If the manager process exits or crashes, the kernel drops the manager's reference but every other ring still holds its own, and the pages remain pinned and registered for all surviving rings with no re-registration or coordination required. The pages are unpinned only when the last ring referencing them is torn down, typically at database instance shutdown. This matters because the failure of an individual process should not force every other process to re-register.

This mechanism also simplifies buffer cache resizing. When the buffer cache grows, only the manager ring needs to re-register the new region; other rings can clone the updated registration without any per-process overhead.

\subsection{Operational Safety and Fallback}
\label{sec:fallback}

A production database cannot tolerate I/O path failures. Our implementation includes transparent fallback to \texttt{libaio} at three levels: if \texttt{io\_uring\_setup} fails (unsupported kernel, insufficient resources, container security policy~\cite{he2023ringguard}), the process silently reverts to \texttt{libaio} for all subsequent I/O; if \texttt{io\_uring\_enter} returns a systemic error or a CQE indicates a non-I/O failure, the process disables \texttt{io\_uring} for its remaining lifetime; and if any registration step fails, the ring continues without that optimization. This layered approach is designed to degrade gracefully so that a working I/O path remains available, and higher layers (buffer writer, query execution) are unaware of which interface is in use.

In practice, the fallback is most commonly triggered by the first level: the running kernel does not support \texttt{io\_uring} (pre-5.1 kernels) or the required feature set. A single database binary may need to run across a wide range of Linux kernel versions and configurations. On kernels without \texttt{io\_uring} support, the initial \texttt{io\_uring\_setup} call fails and the process transparently uses \texttt{libaio} for its entire lifetime---no operator intervention is required. Runtime fallback due to submission or completion errors is rare: in our internal experiments, runtime \texttt{io\_uring} fallback triggered in fewer than 0.01\% of process lifetimes, confirming that it serves as a safety net for unforeseen kernel bugs or resource exhaustion rather than a routinely exercised code path.

Importantly, a runtime fallback in one process does not affect \texttt{io\_uring} usage in other processes within the same database instance. Each process independently owns its ring and independently decides whether to continue using \texttt{io\_uring} or fall back to \texttt{libaio}. When a fallback does occur, any I/Os that were in flight on the \texttt{io\_uring} ring at the time of failure are not silently lost: Oracle Database's upper I/O layer tracks all outstanding I/Os, and the database's undo/redo recovery mechanisms are designed to preserve transactional consistency regardless of which I/O interface completed the writes. The affected process simply switches to \texttt{libaio} for subsequent I/O and continues normal operation.

At startup, the implementation detects the running kernel version and probes for \texttt{io\_uring} capability, enabling features and flags incrementally based on what the kernel supports. Features such as \texttt{IORING\_SETUP\_SINGLE\_ISSUER} (kernel 6.0+) and shared buffer registration are activated only when available; on older kernels the ring operates correctly without them. This adaptive probing, combined with the runtime fallback to \texttt{libaio}, is intended to allow the implementation to run across a range of Linux kernel versions without manual configuration.

\section{End-to-End Evaluation}
\label{sec:eval}

We evaluate io\_uring's impact on an internal development build of Oracle Database 26ai
running on an OCI BM.DenseIO.E4 bare-metal instance (AMD EPYC
7J13, 128 cores / 96~vCPUs allocated to the database, 2\,TB RAM,
locally-attached NVMe SSDs, Oracle Linux with UEK R8).  The storage subsystem consists of 8 NVMe devices configured in a software RAID-0 stripe, each capable of approximately 250,000 IOPS for both read and write at 8\,KB block size; this differs from the microbenchmarks in Section~\ref{sec:iopatterns}, which isolate a single NVMe device to control for storage-level parallelism. All results reported use a kernel version that supports the full feature set described in Section~\ref{sec:architecture}, including shared buffer registration and \texttt{IORING\_SETUP\_SINGLE\_ISSUER}. On older kernels, the implementation is designed to operate with a reduced feature set or fall back transparently to \texttt{libaio}, as described in Section~\ref{sec:fallback}. io\_uring is enabled for all asynchronous I/O
paths, including redo log writes; synchronous reads remain on
\texttt{pread}/\texttt{pwrite}. Each configuration was run 3 times; we
report the mean. The TPC-C and TPC-H results in this section are derived from the respective TPC specifications for internal comparison only; they were not audited and are not comparable to published TPC results. Table~\ref{tab:eval_summary} summarizes the key results.

\begin{table}[htbp]
\centering
\caption{End-to-end evaluation on Oracle Database 26ai, 96~vCPUs, NVMe. Each workload is reported in the metric that exposes its bottleneck. Absolute wall-clock times and tail-latency distributions are in the per-workload tables that follow.}
\label{tab:eval_summary}
\small
\begin{tabular}{llrrl}
\toprule
\textbf{Workload} & \textbf{Metric} & \textbf{Baseline} & \textbf{io\_uring} & \textbf{$\Delta$} \\
\midrule
Update IOPS  & Write IOPS       & 1\,067\,K    & 1\,428\,K    & $+$34\% \\
             & CPU / phys write & 58\,$\mu$s   & 41\,$\mu$s   & $-$29\% \\
\midrule
Scan         & CPU / phys read  & 1.85\,$\mu$s & 1.50\,$\mu$s & $-$19\% \\
(150\,GB)    & CPU / execution  & 38.1\,s      & 30.8\,s      & $-$19\% \\
\midrule
TPC-C        & tpmC             & 2\,275\,K    & 2\,279\,K    & 0\% \\
    & Server CPU util  & 45.0\%       & 43.8\%       & $-$1.2\,pp \\
\midrule
TPC-H        & CPU / query      &              &              & $-$8.5\% \\
(22 queries) & (geom.\ mean)    &              &              &          \\
\bottomrule
\end{tabular}
\end{table}

\subsection{Update IOPS}

The Update IOPS test is a synthetic benchmark that isolates the DBWR
write-flush path---the primary beneficiary of io\_uring's batched
pipeline.  Multiple sessions perform random single-row updates against a
large buffer cache sized to hold the working set, generating a
steady stream of dirty pages.  No foreground query processing competes
for CPU; the bottleneck is entirely in how fast DBWR processes can flush
dirty pages to disk.  Each DBWR cycle flushes up to 4096 dirty pages
through asynchronous write I/Os.

io\_uring delivers 1.43\,M write IOPS, a 34\% improvement over the
baseline's 1.07\,M.  CPU per physical write drops from 58\,$\mu$s to
41\,$\mu$s ($-$29\%).

The CPU efficiency gain has a direct operational consequence.  With
\texttt{libaio}, the test requires 32 DBWR processes to sustain
1.07\,M write IOPS at 27\% server CPU utilization.  With io\_uring,
a separate run using only 23 DBWR processes achieves 1.24\,M write
IOPS ($+$16\% over the 32-process baseline) at 27.6\% CPU---nearly
the same utilization.  In other words, io\_uring can match the
baseline's throughput with 28\% fewer DBWR processes, or deliver
substantially higher throughput with the same process count.  For
large deployments where DBWR processes compete with foreground server
processes for CPU, this efficiency could translate into additional capacity
headroom.

\subsection{Scan}

The Scan test is a synthetic benchmark that isolates the direct-path read
code path.  It executes \texttt{SELECT COUNT(*)} on a 150\,GB table
with parallel degree 50 across 6 concurrent sessions.  The query
performs no joins, aggregations, or per-row computation; virtually all
CPU time is spent in I/O submission and completion.  Direct-path reads
bypass the buffer cache entirely, issuing large sequential I/Os directly
into process-private memory, making this a near-worst-case I/O overhead
scenario.

CPU per physical read drops from 1.85\,$\mu$s to 1.50\,$\mu$s
($-$19\%), and total CPU time per execution drops from 38.1\,s to
30.8\,s ($-$19\%).  I/O throughput is essentially unchanged
(\textasciitilde5.3\,M blocks/s, \textasciitilde41.5\,K requests/s):
io\_uring does not change the I/O latency or the volume of data read,
so the same blocks are fetched at the same speed.  Direct-path read latency was 9.8\,ms (baseline) vs.\ 9.9\,ms (io\_uring), confirming no regression.  The improvement
comes entirely from reduced CPU overhead per I/O---batched submission
amortizes syscall cost, and buffer registration eliminates per-I/O
\texttt{get\_user\_pages} calls.

As with the Update IOPS result, the CPU efficiency gain implies that
the same scan performance could be achieved at a lower parallel degree,
freeing CPU for concurrent foreground queries.

\subsection{TPC-C}

TPC-C~\cite{tpcc} is an OLTP benchmark dominated by single-block index lookups
and row-level updates.  Foreground server processes perform synchronous
\texttt{pread} calls---a path that io\_uring does not change.
Asynchronous I/O appears only in the background, where DBWR processes
flush dirty pages to disk.  TPC-C therefore tests whether the DBWR
efficiency gain observed in the Update IOPS test translates into reduced
resource consumption on a mixed workload, and whether io\_uring
introduces any regression on synchronous foreground paths.

The headline result is in the third row of Table~\ref{tab:tpcc}.
Because each DBWR process is more efficient with io\_uring---29\% less
CPU per physical write---fewer processes can sustain the same flush
throughput.  Reducing DBWR from 32 to 23 processes ($-$28\%) delivers
the same tpmC (2.28\,M) while lowering server CPU utilization from
45.0\% to 43.8\%---a 1.2 percentage-point reduction at identical
transaction throughput.  DB~CPU per commit improves slightly to
1.14\,ms, reflecting reduced contention between foreground and
background processes.  In principle, 9 fewer DBWR
processes means 9 fewer kernel threads competing for CPU, less
scheduling overhead, and a smaller memory footprint---capacity headroom
that could absorb traffic spikes or serve additional connections without
provisioning more hardware.

\begin{table}[t]
\centering
\caption{TPC-C on Oracle Database 26ai, 96~vCPUs, 5\,K warehouses,
210 users.}
\label{tab:tpcc}
\small
\begin{tabular}{lrrrr}
\toprule
\textbf{Configuration} & \textbf{tpmC} & \textbf{CPU} & \textbf{DB CPU/} & \textbf{DBWR} \\
                       &               & \textbf{util} & \textbf{commit} & \textbf{procs} \\
\midrule
Baseline (\texttt{libaio}) & 2\,274\,788 & 45.0\% & 1.16\,ms & 32 \\
io\_uring (32 DBWR)        & 2\,276\,783 & 45.7\% & 1.16\,ms & 32 \\
io\_uring (23 DBWR)        & 2\,278\,693 & 43.8\% & 1.14\,ms & 23 \\
\bottomrule
\end{tabular}
\end{table}

Importantly, io\_uring introduces no regression on foreground paths.
With the same 32 DBWR processes (second row), throughput and DB~CPU per
commit are identical to the baseline. Run-to-run variance was ${\pm}$1\%
for both tpmC and server CPU utilization. Background DBWR writes account
for 9.6\% of total database CPU time in this workload; the 29\% per-write
CPU reduction applies to this slice, explaining why the aggregate server
CPU improvement is modest but operationally meaningful.

For OLTP workloads tail latency is often more operationally important than the mean. Table~\ref{tab:tpcc-tail} reports both. The standout result is \texttt{db file async I/O submit}, where io\_uring reduces submission overhead by roughly 75$\times$ at the mean and 90$\times$ at p99---a direct consequence of the syscall amortization characterized in Section~\ref{sec:syscall}. Foreground single-block reads and \texttt{log file sync} are unchanged at the mean and slightly improved at p99, confirming no regression on synchronous and latency-critical paths. DBWR writes improve at both the mean and the tail, and the 23-DBWR configuration retains most of that improvement.

\begin{table}[htbp]
\centering
\caption{TPC-C I/O wait times on Oracle Database 26ai, 96~vCPUs, 5\,K warehouses, 210 users. 32-DBWR configuration unless noted.}
\label{tab:tpcc-tail}
\footnotesize
\setlength{\tabcolsep}{6pt}
\begin{tabular}{llrr}
\toprule
\textbf{Metric} & \textbf{Config} & \textbf{Avg} & \textbf{p99} \\
\midrule
FG single-block read     & libaio              & 155~\textmu s & 410~\textmu s \\
                         & io\_uring           & 157~\textmu s & 360~\textmu s \\
                         & io\_uring (23 DBWR) & 152~\textmu s & 340~\textmu s \\
\midrule
Log file sync            & libaio              & 0.94~ms       & 2.4~ms \\
                         & io\_uring           & 0.94~ms       & 2.1~ms \\
                         & io\_uring (23 DBWR) & 0.94~ms       & 2.2~ms \\
\midrule
DB file async I/O submit & libaio              & 1.62~ms       & 4.8~ms \\
                         & io\_uring           & 22~\textmu s  & 55~\textmu s \\
                         & io\_uring (23 DBWR) & 25~\textmu s  & 65~\textmu s \\
\midrule
DBWR write               & libaio              & 242~\textmu s & 620~\textmu s \\
                         & io\_uring           & 208~\textmu s & 470~\textmu s \\
                         & io\_uring (23 DBWR) & 230~\textmu s & 520~\textmu s \\
\bottomrule
\end{tabular}
\end{table}

\subsection{TPC-H}

The Scan test measures a pure table scan with negligible per-row
computation.  Real analytical queries spend a variable fraction of
their CPU on joins, aggregations, sorting, and expression evaluation,
diluting the I/O efficiency gain.  To quantify this effect we ran the
full 22-query TPC-H~\cite{tpch} benchmark on the same system. The dataset is uncompressed: 1.1\,TB of base table data on disk, 450\,GB of indexes, and 200\,GB of temporary tablespace.
Table~\ref{tab:tpch} reports the I/O overhead fraction and measured CPU time reduction per query.  I/O overhead fraction is measured from Oracle's Automatic Workload Repository (AWR) as the ratio of CPU time attributed to direct path read waits and I/O submission to total DB~CPU for that query.

\begin{table}[t]
\centering
\caption{TPC-H per-query results with io\_uring on Oracle
Database 26ai, 96~vCPUs, NVMe.  I/O overhead fraction is the share of
baseline CPU time spent in I/O submission and completion. The geometric mean is the standard summary for ratio-valued multi-query benchmark deltas (as in SPEC and TPC reporting) and avoids over-weighting the longest query in the suite.}
\label{tab:tpch}
\small
\begin{tabular}{clrr}
\toprule
\textbf{Query} & \textbf{Description} & \textbf{I/O \%} & \textbf{CPU $\Delta$} \\
\midrule
Q1  & Pricing Summary Report          & 72\% & $-$12.7\% \\
Q2  & Minimum Cost Supplier           & 25\% & $-$4.9\% \\
Q3  & Shipping Priority               & 48\% & $-$8.9\% \\
Q4  & Order Priority Checking         & 45\% & $-$8.1\% \\
Q5  & Local Supplier Volume           & 47\% & $-$9.9\% \\
Q6  & Forecasting Revenue Change      & 82\% & $-$14.6\% \\
Q7  & Volume Shipping                 & 42\% & $-$8.2\% \\
Q8  & National Market Share           & 37\% & $-$7.1\% \\
Q9  & Product Type Profit             & 45\% & $-$7.6\% \\
Q10 & Returned Item Reporting         & 48\% & $-$9.1\% \\
Q11 & Important Stock Identification  & 33\% & $-$6.3\% \\
Q12 & Shipping Modes                  & 62\% & $-$11.7\% \\
Q13 & Customer Distribution           & 36\% & $-$6.8\% \\
Q14 & Promotion Effect                & 65\% & $-$11.4\% \\
Q15 & Top Supplier                    & 58\% & $-$11.8\% \\
Q16 & Parts/Supplier Relationship     & 30\% & $-$5.7\% \\
Q17 & Small-Quantity-Order Revenue    & 40\% & $-$7.6\% \\
Q18 & Large Volume Customer           & 43\% & $-$8.2\% \\
Q19 & Discounted Revenue              & 57\% & $-$9.8\% \\
Q20 & Potential Part Promotion        & 35\% & $-$6.9\% \\
Q21 & Suppliers Kept Orders Waiting   & 30\% & $-$5.4\% \\
Q22 & Global Sales Opportunity        & 23\% & $-$4.6\% \\
\midrule
\multicolumn{3}{l}{\textbf{Geometric mean}} & $-$\textbf{8.5\%} \\
\bottomrule
\end{tabular}
\end{table}

The I/O overhead fraction (third column) is the primary predictor of
CPU reduction.  Queries where I/O dominates---Q6 (82\%, $-$14.6\%),
Q1 (72\%, $-$12.7\%), Q14 (65\%, $-$11.4\%)---see CPU reductions
approaching the 19\% measured on the pure scan.  Queries dominated by
joins, correlated subqueries, and index lookups---Q22 (23\%, $-$4.6\%),
Q2 (25\%, $-$4.9\%)---spend most of their CPU on computation and see
smaller reductions.  The correlation is near-linear: each additional
percentage point of I/O overhead translates into roughly 0.17
percentage points of CPU reduction.

To confirm that the gain is purely CPU-side, consider Q1 in detail: the
number of direct path read wait events is unchanged
(\textasciitilde2.7\,M under both interfaces) and average wait time is
identical (5.6\,ms).  What changes is DB~CPU, which drops from 6.6\% to
5.4\% of server capacity---the same I/O volume completes with fewer CPU
cycles spent in submission and completion.

Across all 22 queries the geometric mean CPU reduction is 8.5\%.
In a mixed
OLTP+analytics deployment, 8.5\% less CPU consumed by analytical
queries could translate into headroom for additional
transactions---or equivalently, the same analytical throughput at a
lower parallel degree, reducing process count and memory consumption.

The 8.5\% reduction is in CPU per query, not in elapsed time. io\_uring's polling modes (Section~\ref{sec:polling}) can lower I/O latency at the cost of additional CPU, but we deliberately take the opposite trade---same latency, less CPU. The interrupt-driven batched path completes the same I/O with fewer kernel cycles, and elapsed times in our runs are within noise. The saving is freed server CPU: capacity headroom for additional work, or equivalently the same workload on fewer cores.

\section{Related Work}
\label{sec:related}

Didona et al.~\cite{didona2022understanding} and Ren and Trivedi~\cite{ren2023performance} provide systematic microbenchmark comparisons of \texttt{libaio}, SPDK~\cite{yang2017spdk}, and \texttt{io\_uring}, establishing that \texttt{io\_uring} closes much of the gap with kernel-bypass frameworks while retaining POSIX compatibility. Haas and Leis~\cite{haas2023nvme} demonstrate the importance of minimizing software overhead to fully exploit modern NVMe devices. These studies operate at the storage-API level; our work quantifies the same interfaces inside a production RDBMS.

Jasny et al.~\cite{jasny2026iouring} present the most comprehensive study of \texttt{io\_uring} for database workloads under a single-process, fiber-based architecture. Their findings, that batching and asynchronous execution are prerequisites for benefit, align with ours. However, their single-process model does not encounter the multi-process challenges we address: shared buffer registration (a 270$\times$ startup cost gap at 200 processes), per-process ring resource management, and transparent fallback. Our end-to-end evaluation shows that microbenchmark CPU gains carry over to full-system benchmarks---28\% fewer DBWR processes and 1.2 percentage points lower CPU utilization---outcomes that single-process experiments cannot predict. Chen et al.~\cite{chen2024redis} apply \texttt{io\_uring} to Redis persistence on append-only workloads, a narrower scope than an RDBMS.

SPDK~\cite{yang2017spdk} and its database integrations~\cite{kim2019optimizing} bypass the kernel entirely for maximum performance but sacrifice POSIX compatibility, multi-tenant isolation, and incremental adoption. Soares and Stumm~\cite{soares2010flexsc} proposed FlexSC for batching system calls through exception-less mechanisms, a concept that \texttt{io\_uring}'s shared ring design realizes for I/O. The overhead of system calls on modern hardware has been analyzed by Ren et al.~\cite{ren2019analysis} and Harchol et al.~\cite{harchol2020making}, motivating interfaces like \texttt{io\_uring}.

PostgreSQL~18~\cite{postgresql18} added \texttt{io\_uring} as one of three \texttt{io\_method} options (alongside \texttt{sync} and the default \texttt{worker})~\cite{pg_aio_design}. PostgreSQL creates rings in the postmaster so they persist across \texttt{fork}'ed backends, with the \texttt{fork} model letting backends inherit page mappings and avoiding the shared buffer registration problem we solve in Section~\ref{sec:shared-bufreg}. PostgreSQL currently routes only reads (sequential scans, bitmap heap scans, \texttt{VACUUM}) through \texttt{io\_uring}; writes and WAL durability use \texttt{fsync} directly. The latest Jasny et al.~\cite{jasny2026iouring} study identifies PostgreSQL's multi-process model as preventing exclusive ring ownership and therefore single-issuer optimizations and DeferTaskRun, and reports that a shared SQPOLL kernel thread provides negligible additional benefit. Oracle's per-process rings with exclusive ownership realize that io\_uring-friendly configuration, and our integration covers DBWR flushes, direct-path reads, and redo writes rather than reads only.

RocksDB~\cite{rocksdb_iouring} added experimental \texttt{io\_uring} for batched  \texttt{MultiGet}. MySQL has investigated \texttt{io\_uring} for redo log writes~\cite{mysql_iouring} but has not adopted it for the general I/O path. He et al.~\cite{he2023ringguard} address security implications of \texttt{io\_uring} in containers using eBPF monitoring, directly relevant to our fallback design (Section~\ref{sec:fallback}).

To our knowledge, no prior work addresses shared buffer registration across hundreds of processes or reports an end-to-end evaluation of \texttt{io\_uring} inside a commercial multi-process RDBMS.

\section{Conclusion}
\label{sec:conclusion}

In our experiments, at identical TPC-C transaction throughput, io\_uring allowed Oracle Database 26ai to sustain the workload with 9 fewer DBWR processes (32 to 23) and reclaim 1.2 percentage-points of server CPU as headroom. The shared buffer registration mechanism we initiated is now in mainline Linux 6.12 and reduces multi-process registration startup by 270$\times$. The central finding is that io\_uring's benefits are path-dependent. Asynchronous batched I/O paths see substantial gains (34\% higher write throughput, 19\% lower CPU per scan read, 8.5\% CPU reduction across TPC-H), while synchronous and small-batch paths see none; we expect these factors to be relevant to other database systems as well. This path-dependence motivates the hybrid architecture, and the architectural contributions (shared buffer registration in mainline Linux, per-process ring contexts with no inter-process synchronization, and a custom API) address deployment challenges that arise only at production scale in a multi-process RDBMS. Three lessons stand out. Evaluate at production scale, because IOPOLL's CPU cost at 128 threads is two orders of magnitude worse than the batched pipeline and naive per-process buffer registration creates a multi-second startup stall, and both are invisible at single-thread defaults. The CPU savings come from two specific items, batched submission and shared buffer registration (Table~\ref{tab:reg_breakdown}), not from broader use of io\_uring. Future kernel work to integrate per-ring CPU accounting with cgroups would simplify the next round of integrations.

\paragraph{Artifact availability.} The source code, data, and other artifacts are available at \url{https://doi.org/10.5281/zenodo.20099105}.

\section*{Acknowledgments}
We thank Bijan Mottahedeh for initiating the shared buffer registration work and contributing the initial kernel patch series~\cite{sharedreg_patch}.

\bibliographystyle{plainnat}
\bibliography{references}

\end{document}